\documentclass[preprint,aps,showpacs,11pt]{revtex4-1}
\usepackage{hyperref}
\usepackage{amsmath}
\usepackage{amsfonts}
\usepackage{amssymb}
\usepackage{graphicx}
\def\gsim{\lower.7ex\hbox{$\;\stackrel{\textstyle>}{\sim}\;$}}
\def\lsim{\lower.7ex\hbox{$\;\stackrel{\textstyle<}{\sim}\;$}}
\begin{document}
\title{Couplings of quarks in the Partially Aligned 2HDM with a four-zero texture Yukawa matrix}
\author{J. Hern\'andez-S\'anchez}
\email{jaimeh@ece.buap.mx}
\affiliation{Fac. de Cs. de la
Electr\'onica, Benem\'erita Universidad Aut\'onoma de Puebla, Apdo. Postal 542, C. P. 72570 Puebla, Puebla, M\'exico, and Dual C-P Institute of High Energy Physics, M\'exico.}
\author{L. L\'opez-Lozano}
\affiliation{Fac. de Cs. de la
Electr\'onica, Benem\'erita Universidad Aut\'onoma de Puebla, Apdo. Postal 542, C. P. 72570 Puebla, Puebla, M\'exico.}
\author{R. Noriega-Papaqui}
\email{rnoriega@uaeh.edu.mx}
\affiliation{\'Area Acad\'emica de Matem\'aticas y F\'{\i}sica, Universidad Aut\'onoma del Estado
de Hidalgo, Carr. Pachuca-Tulancingo Km. 4.5, C.P. 42184, Pachuca, Hgo and Dual C-P Institute of High Energy Physics, M\'exico.}
\author{A. Rosado}
\email{rosado@sirio.ifuap.buap.mx} \affiliation{Instituto de F\'isica, BUAP. Apdo. Postal J-48, C.P. 72570 Puebla, Pue., M\'exico}
\date{\today}
\begin{abstract}
The Two Higgs Doublets Model (2HDM) has provided a very useful way
to describe a minimal extension of the scalar sector of the Standard
Model. In this work, it is shown a scheme that we call Partial 
Aligned Two Higgs Doublet Model (PA-2HDM) which allows a description
of the distinct versions of the 2HDM in a simple way, including
those with flavor symmetries. In addition, it is shown a method to
diagonalize Yukawa matrices of four-zero texture coming from the
2HDM-III. We provide some phenomenological applications  in order to show the 
model's predictive power.
\end{abstract}
\pacs{12.15.-y,12.60.-i,12.60.Fr}
\maketitle

The main problem in the flavor physics beyond the Standard Model (SM) \cite{stanmod} is to control the presence of the Flavor Changing Neutral Currents (FCNC) that in experiments have been
observed to be highly suppressed. Almost all models that describe physics in energy regions greater than the electroweak scale, have contributions with FCNC at tree level, unless some symmetry
is introduced on the scalar sector to suppress them. One of the most important extensions of the SM is the Two Higgs Doublet Model (2HDM), due to its wide variety of dynamical features and the
fact that it can represent a low-energy limit of general models like the Minimal Supersymmetric Standard Model. There are some generalizations of the 2HDMs of type I, II, X and Y (2HDM-I,
2HDM-II, 2HDM-X and 2HDM-Y) \cite{Aoki:2009ha}, as well as the 2HDM-III with flavor symmetries that require a four texture in the Yukawa matrix \cite{DiazCruz:2004pj} and Lepton Flavor
Violating (LFV) introduced as a deviation from Model II Yukawa interaction \cite{Kanemura:2005hr,Kanemura:2004cn}. The type-X (type-Y) 2HDM is referred to as the type-IV (type-III) 2HDM in
Ref.\cite{Barger:1989fj} and  the type-I' (type-II') 2HDM in Ref. \cite{Grossman:1994jb,hep-ph/9603445}. Sometimes, the most general 2HDM, in which each fermion couples to both Higgs doublet
fields, is called the type III 2HDM \cite{Liu:1987ng}. From a phenomenological point of view, the Cheng-Sher {\it ansatz} \cite{Cheng:1987rs} has been very useful to describe the
phenomenological content of the Yukawa matrix and the salient feature of the hierarchy of quark masses. Through the Yukawa textures \cite{fritzsch,fourtext} it is possible to build a matrix
that preserves the expected Yukawa couplings that depend on the fermion masses. One unavoidable problem is the great number of free parameters that emerge as a consequence of introducing a new
Higgs doublet. In order to reduce the number of free parameters of the model some restrictions have been imposed on the entries of the Yukawa matrices through discrete symmetries or phenomenological
assumptions. The Yukawa Lagrangian for the quark fields is given by
\begin{equation}\label{2HDMlagrangian}
\mathcal{L}_Y = Y_1^u\bar{Q}_L \tilde{\Phi}_1u_R + Y_2^u\bar{Q}_L\tilde{\Phi}_2 u_R+Y_1^d\bar{Q}_L
\Phi_1d_R+Y_2^d\bar{Q}_L\Phi_2d_R
\end{equation}
where $\Phi_{1,2}=(\phi^+_{1,2},\phi^0_{1,2})^T$ denotes the Higgs doublets,
 $\tilde{\Phi}_{1,2} = i \sigma_2 \Phi_{1,2}^*$ and $Y^{u,d}$ are the Yukawa matrices.

The above Lagrangian (\ref{2HDMlagrangian}) has a great deal of free
parameters associated with the Yukawa interaction and five scalar
bosons, two of them charged ($H^\pm$) and one of the neutral ones is a
pseudoscalar ($A^0$). The mechanism through which the FCNC are
controlled defines the version of the model and a different
phenomenology that can be contrasted with the experiment. One
successful version where the Yukawa couplings depend on the
hierarchy of masses is the one where the mass matrix has a four-zero
texture form. This matrix is based on the phenomenological
observation that the off-diagonal elements must be small in order to
dim the interactions that violate flavor as the experimental results
show. Although the phenomenology of Yukawa couplings constrains the
hierarchy of the mass matrix entries, it is not enough to determine
the strength of the interaction with scalars. Another assumption on
the Yukawa matrix is related to the additional Higgs doublet. In
versions I and II  it is introduced a discrete symmetry on the Higgs
doublets, fulfilled by the scalar potential, that leads to the
vanishing of most of the free parameters. However, version III,
having a richer phenomenology, requires a slightly more general
scheme.

There is a close relation between the flavor space and the mass
matrix, which in general can be written as
\begin{equation}
M_f = \frac{1}{\sqrt{2}}(v_1Y_1^f+v_2Y_2^f).
\end{equation}
Inspired by the fact that in the Higgs basis the information of the mass matrix is contained in the first Yukawa matrix and the SM couplings are proportional to the fermion masses, the
interactions with scalars in a general 2HDM can be modeled by imposing a specific form on the second Yukawa matrix as a mass matrix transformed in the flavor space. In this paper it is utilized
a particular case of this model \cite{OurModel} that can describe different versions of the 2HDM by using properties of the flavor space through a simple principle. We introduce the concept
Partially Aligned (PA) Yukawa Matrix according to two criteria: a) a new transformation  for the first Yukawa matrix in the flavor space $SU_F(3)$ and b) the control of FCNC induced by this
transformation, using as a criterion the Cheng-Sher  {\it ansatz}  \cite{Cheng:1987rs}. By following these ideas, the concept of Partially Aligned (PA)  will be defined by a new transformation
which enables us to write the matrix of couplings as a bi-unitary transformed mass matrix, namely
\begin{equation}
Y^f_2 = \frac{1}{v}A^f_LM_fA^f_R,
\end{equation}
where $A^f_L$ and $A^f_R$ with $f=u,d,\ell$, are diagonal $SU_F(3)$
matrices that concentrate the dynamical information about extended
scalar interactions and $M_f$ contains the properties of the
hierarchy of the quark masses and the mixing of the CKM matrix,
whose form is determined by a more fundamental theory. As usual, we have combined the VEVs of the doublet Higgs
fields through the relation $v^2=v^2_1+v^2_2$.
In the
PA-2HDM the aligned model \cite{Pich:2009sp,Zhou:2003kd}
can be cast with $A^f_L=A^f_R\sim \lambda_0$, where $\lambda_0$ is
the matrix proportional to a unit matrix in $SU_F(3)$. Details about
these formulations are given elsewhere. As mentioned above, the
several versions of the 2HDM can be generated by choosing suitable
matrices (see table \ref{versions2HDM}). There is no physical restriction on the structure of the mass
matrix beyond the fact that the quark masses of different families
differ by several orders of magnitude.
\begin{center}
\begin{table}
\begin{tabular}{ccccc}
\hline
\hline
&$A_L^ u$&$A_R^u$&$A_L^d$&$A_R^d$\\
\hline
I&$\sqrt{\frac{3m_W}{v}}\lambda_0 $&$\sqrt{\frac{3m_W}{v}}\lambda_0$&$
\sqrt{\frac{3m_W}{v}}\lambda_0$&$\sqrt{\frac{3m_W}{v}}\lambda_0$\\
II&$\sqrt{\frac{3m_W}{v}}\lambda_0$&$\sqrt{\frac{3m_W}{v}}\lambda_0$&$0_{3\times 3}$&$0_{3\times 3}$\\
III-IV&$\sum_{a=0,3,8} C^u_{a}\lambda_a$&$\left(\sum_{a=0,3,8} \tilde{C}^u_a\lambda_a\right)^\dagger$&$\sum_{a=0,3,8}C^d_a\lambda_a$&$\left(\sum_{a=0,3,8}\tilde{C}^d_a\lambda_a\right)^\dagger$\\
A2HDM&$C_0^u\lambda_0$&$\tilde{C}_{0}^{u*}\lambda_0$&$C_0^d\lambda_0$&$\tilde{C}_{0}^{d*}\lambda_0$\\
\hline
\hline
\end{tabular}
\caption{Matrices that reproduce several versions of the Yukawa interactions
for the 2HDM in terms of $SU_F(3)$ generators. The $C's$ parameters are complex coefficients and they are proportional to the parameters
$\widetilde{\chi}_{ij}^f$ defined in Eq.(\ref{def-chi}).}
\label{versions2HDM}
\end{table}
\end{center}
On the other hand, the PA-2HDM will induce Higgs boson FCNC  through the following  term
\begin{equation}
\label{eq1}
\widetilde{Y}_2^f = \frac{1}{v} \widetilde{A}^f_L \bar{M}_f \widetilde{A}^f_R,
\end{equation}
where $\widetilde{A}^f_{L,R} = U_{L,R}^{f \dagger} A^f_{L,R} U_{L,R}^f$, $\bar{M}_f = \textrm{Diag}\lbrack m_{f1}, m_{f2}, m_{f3} \rbrack$ and $U_{L,R}^f$ are the matrices that diagonalize the
mass matrix $M_f$. So,  the contribution to fermion-fermion-Higgs bosons couplings is given by:
\begin{eqnarray}
(\widetilde{Y}_{2}^f )_{ij}= \frac{1}{v} \left( m_{f1} (\widetilde{A}_{L}^f)_{i1} (\widetilde{A}_{R}^f)_{1j} +
m_{f2} (\widetilde{A}_{L}^f)_{i2} (\widetilde{A}_{R}^f)_{2j} +
m_{f3} (\widetilde{A}_{L}^f)_{i3} (\widetilde{A}_{R}^f)_{3j}
\right).
\label{gen-Y}
\end{eqnarray}
In order to control the FCNC induced by the model, we employ  the Cheng-Sher \textit{ansatz} \cite{Cheng:1987rs} in the following way:
\begin{eqnarray}
\label{eq2}
(\widetilde{Y}_{2}^{CS,f})_{ij}=\frac{\sqrt{m_{fi} m_{fj}}}{v} \widetilde{\chi}_{ij}^f,
\label{def-chi}
\end{eqnarray}
then, from  Eq.(\ref{eq1}) and Eq.(\ref{eq2}) the FCNC will be controlled by:
\begin{eqnarray}
 \left| m_{f1} (\widetilde{A}_{L}^f)_{i1} (\widetilde{A}_{R}^f)_{1j} +
m_{f2} (\widetilde{A}_{L}^f)_{i2} (\widetilde{A}_{R}^f)_{2j} + m_{f3} (\widetilde{A}_{L}^f)_{i3} (\widetilde{A}_{R}^f)_{3j} \right| \leq \sqrt{m_{fi} m_{fj}} \left| \widetilde{\chi}_{ij}^f
\right|. \label{gen-chi}
\end{eqnarray}
The advantage of this criterion is that we can use previous studies of the experimental constraints imposed on the free parameters of Yukawa texture \cite{DiazCruz:2004pj,DiazCruz:2004tr,GomezBock:2005hc,DiazCruz:2009ek,BarradasGuevara:2010xs}.
Moreover, by definition, the eigenvalues of the mass matrix are the masses of
fermions, {\it i.e.}, they must be real and non-negative. A
hermitian matrix guarantees that the masses are real, however, the
non-negativity condition for the eigenvalues is not fulfilled by any
hermitian matrix. Actually, the four-zero texture matrix has at
least one negative eigenvalue. This drawback is solved by
considering that upon diagonalization the masses are the square root
of the eigenvalues of $H_f= M_f M_f^\dagger$. This assumption is
appropriate to determine the phenomenological couplings in the
Yukawa sector as the Cheng-Sher {\it ansatz}, albeit, 
it leaves out other possible parameterizations emerging from a
rather general method.

\begin{figure}
\centering
\includegraphics[scale=0.75]{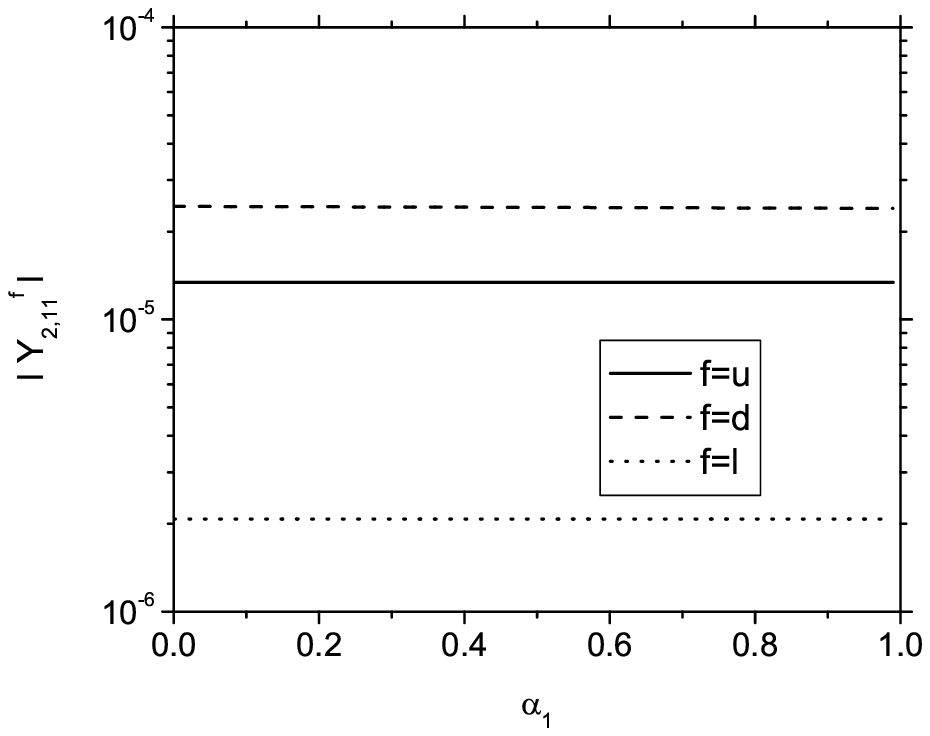}
\includegraphics[scale=0.75]{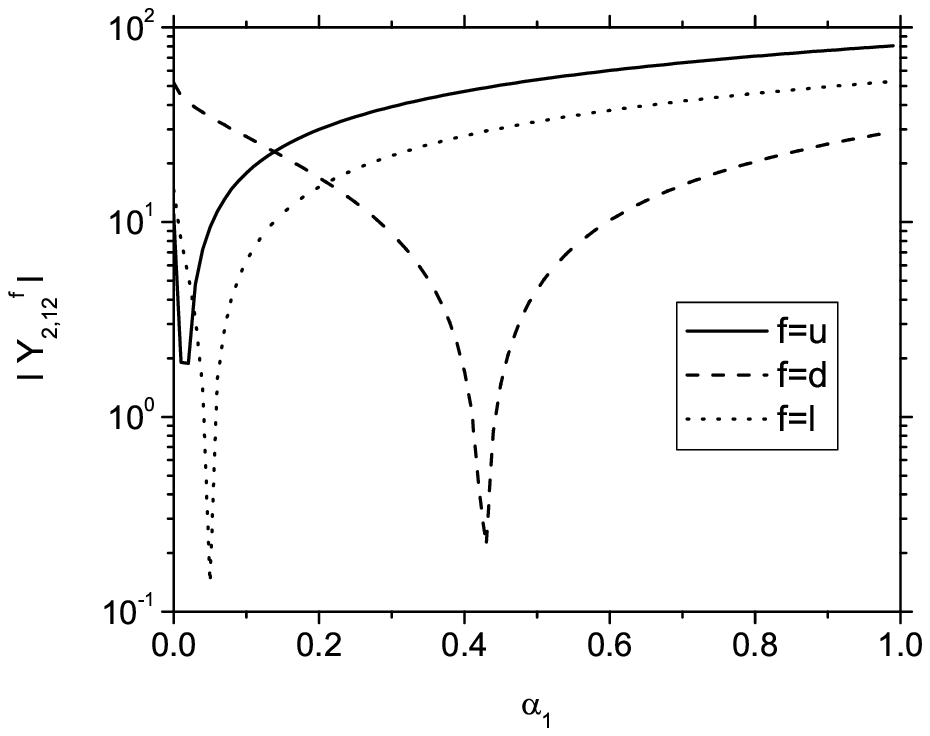}
\includegraphics[scale=0.75]{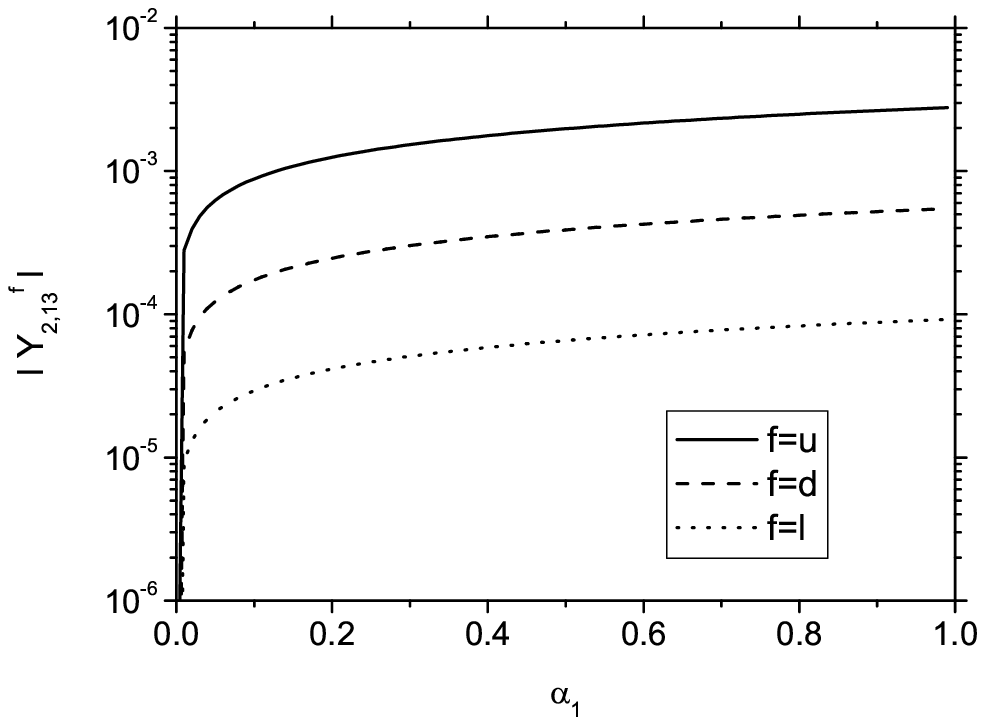}
\includegraphics[scale=0.75]{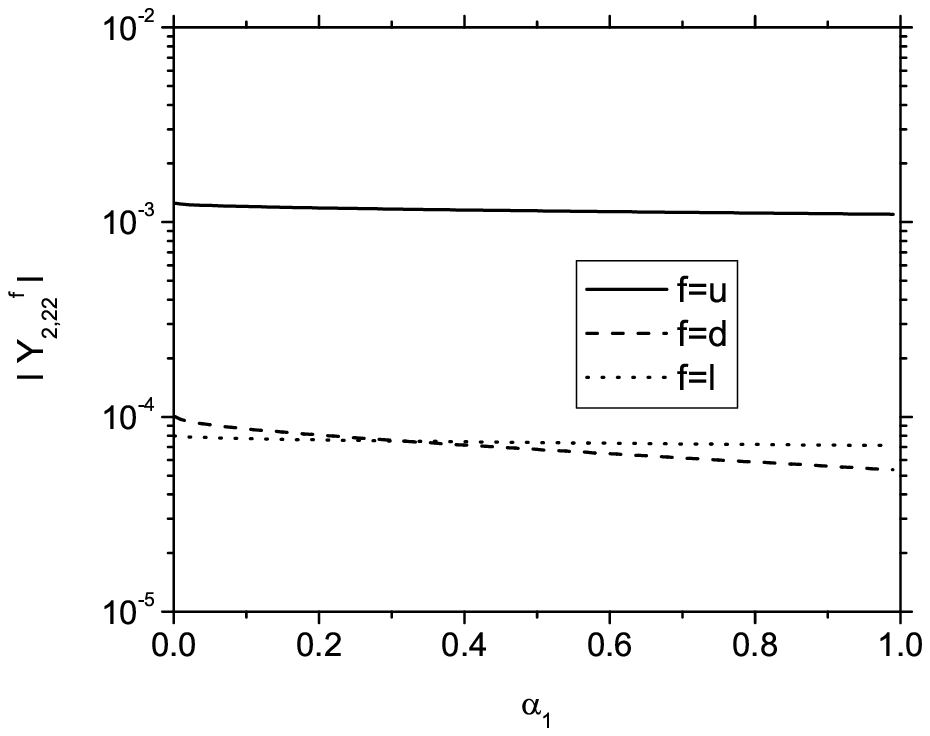}
\includegraphics[scale=0.75]{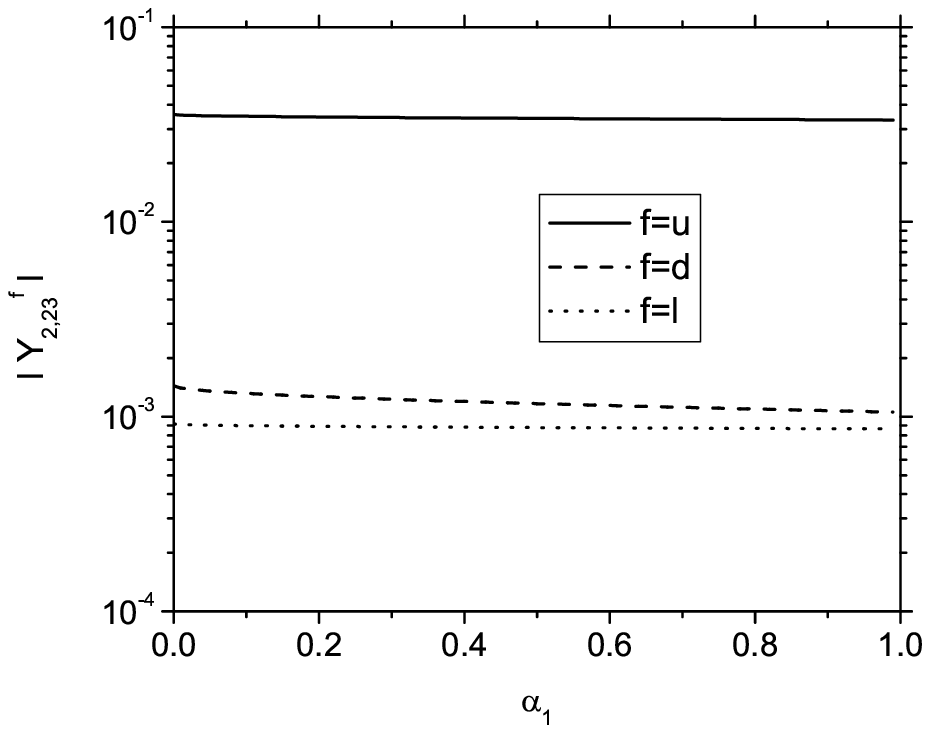}
\includegraphics[scale=0.75]{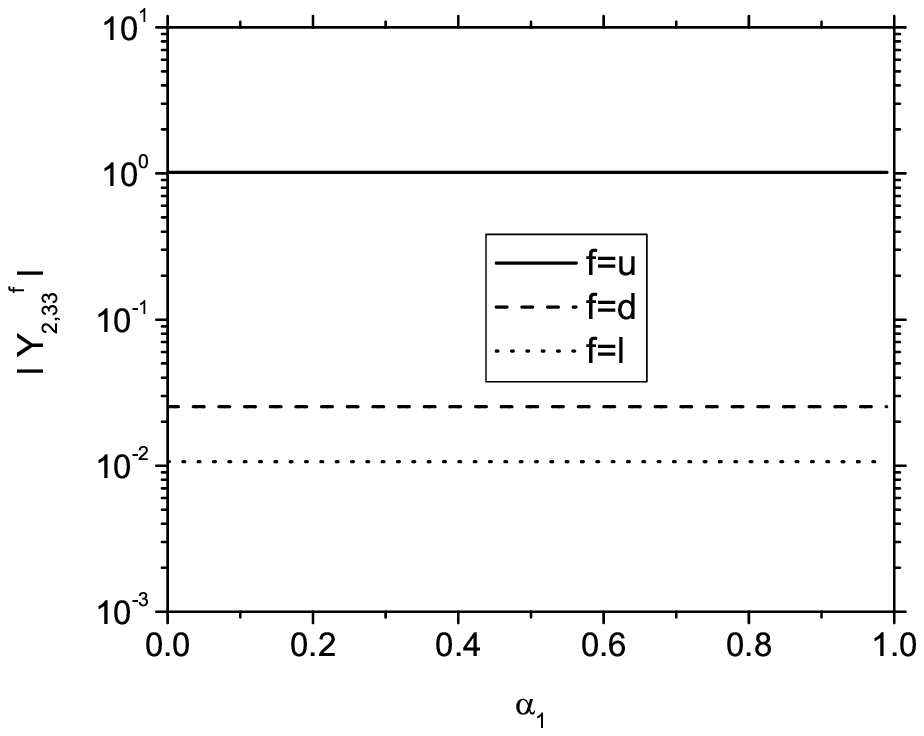}
\caption{Magnitude of $|Y_{2,ij}^f|$ in the limit when the
Cheng-Sher couplings are $|\widetilde{\chi}_{ij}|\sim 1$ and the
phases are taken to be zero} \label{yukawa}
\end{figure}

In the following, instead of making assumptions on the nature of the
eigenvalues of $M_f$, we look for the properties of free parameters
of the mass matrix in the flavor basis in order to generate real and
non-negative eigenvalues of $M_f$. In general, the bi-unitary
transformation is given by $M_f= U^f_L M_f U_R^{f \dagger}$, where $U_L^f$ and
$U_R^f$ represent the unitary transformations that diagonalize
$M_f M_f^\dagger$ and $M_f^\dagger M_f$, respectively. If $M_f$ were
hermitian then $U_L^f$ and $U_R^f$ would be equal. In what follows we
shall restrict ourselves to this case. The most suitable mass matrix
structure that describes the properties of the couplings of the
Yukawa sector is the four-zero texture Yukawa matrix, which can be
written as \cite{DiazCruz:2004tr,DiazCruz:2004pj}
\begin{equation}
M_f=\left(
\begin{array}{ccc}
 0 & C_f & 0 \\
 C_f^* & \widetilde{B}_f& B_f \\
 0 & B_f^* & A_f
\end{array}
\right).
\end{equation}
The form and hierarchy of the free parameters are conserved, although, for the sake of 
generalization, we slightly break the hermitian condition to
allow for new possible effects coming from relative phases between
diagonal elements $A_f$ and $\tilde{B}_f$. It is  worth mentioning that
these phases must obey the physical condition imposed on the
eigenvalues of the mass matrix,
\begin{eqnarray}
A_f &=& |A_f|\cdot e^{i \theta_{Af} },\\
\widetilde{B}_f &=& |\widetilde{B}_f|\cdot e^{i \theta_{\widetilde{B}f}}.
\end{eqnarray}
The hermitian matrix $H_f$ is given by
\begin{equation}
H_f = \begin{pmatrix}
          |C_f|^2   &   \widetilde{B}_f^*C_f   &   B_f C_f   \\
          \widetilde{B}_f C_f^*   &   |C_f|^2 + |\widetilde{B}_f|^2 + |B_f|^2   &   \widetilde{B}_f B_f +A^*_f B_f    \\
          B_f^*C_f^*   &  \widetilde{B}_f^*B^*_f + A_f B_f^*   &   |A_f|^2 + |B_f|^2   \\
       \end{pmatrix}
\label{matrizH}.
\end{equation}
As usual, we can extract the phases of the non-diagonal elements
with a transformation  $H_f = P_f^\dagger \overline{H}_f P_f$ where
$P_f = \textrm{Diag}\lbrack
1, e^{-i(\theta_{\widetilde{B}f}-\theta_{Cf})}, e^{i(\theta_{Bf} + \theta_{Cf})} \rbrack$.
It is important to highlight the unitary relation between $H_r$ and
the diagonal matrix $M_f^2= \textrm{Diag}\lbrack m_{f1}^2,m_{f2}^2,m_{f3}^2 \rbrack$ that
leads to the following system of equations:
\begin{widetext}
\begin{eqnarray}
&& |A_f|^2 + 2|B_f|^2 + 2|C_f|^2 + |\widetilde{B}_f|^2 = m_{f1}^2 + m_{f2}^2 + m_{f3}^2, \label{inv1a}\\
&& (|B_f|^2 + |C_f|^2)^2 + |A_f|^2 |\widetilde{B}_f|^2 + 2|A_f|^2|C_f|^2 -2\cos(\theta_{Af} + \theta_{\widetilde{B}f})
|A_f||\widetilde{B}_f| |B_f|^2   \nonumber\\
&& = m_{f1}^2 \, m_{f2}^2 + m_{f1}^2 \, m_{f3}^2 + m_{f2}^2 \, m_{f3}^2,\label{inv1b}\\
&& |A_f|^2 |C_f|^4 = m_{f1}^2 \, m_{f2}^2 \, m_{f3}^2.\label{inv1c}
\end{eqnarray}
\end{widetext}
Thus, the problem is reduced to solving the system of equations,
given by Eqs.(\ref{inv1a}-\ref{inv1c}), which has 16 possible
solutions, though most of them unphysical. However, by adopting a simple criterion we can to simplify this system of
equations. As mentioned above, we are only interested in those
solutions which reproduce real and non-negative eigenvalues. To
achieve this goal we establish the following three conditions that
ensure physical properties of the solutions:
\begin{itemize}
\item The free complex parameters $A_f$, $B_f$, $C_f$ and $\widetilde{B}_f$ are a function
of the masses and satisfy the invariant equations
(\ref{inv1a}-\ref{inv1c}).
\item The eigenvalues of $M_f$, namely, the fermion masses, must be real and non-negative.
\item The eigenvalues of $H_f$ must obey the hierarchy of the quark masses as experimentally observed,
{\it i.e.}, $m_{f3} > m_{f2} > m_{f1}$ \cite{Nakamura:2010zzi}.
\end{itemize}
A more simplified system of equations is thus obtained by
factorization due to the chosen phases which must fulfill the above conditions:
\begin{widetext}
\begin{eqnarray}
|A_f| + (-1)^m|\widetilde{B}_f| &=& m_{f1} -m_{f2} + m_{f3} \label{inv2a},\\
|B_f|^2 -(-1)^m|A_f||\widetilde{B}_f| + |C_f|^2 &=& m_{f1} \, m_{f2} - m_{f1} \, m_{f3} + m_{f2}\, m_{f3} \label{inv2b},\\
|A_f||C_f|^2 &=& m_{f1} \, m_{f2} \, m_{f3}.
\end{eqnarray}
\end{widetext}
with $m$ integer.
The solution is given by
\begin{eqnarray}
|B_f| &=& \sqrt{\left(1-\frac{m_{f1}}{|A_f|}\right)(|A_f| + m_{f2})(m_{f3}-|A_f|)}  \label{parametro1c1},\\
|\widetilde{B}_f| &=& (-1)^m (m_{f1} -m_{f2} +m_{f3}-|A_f|) \label{parametro2c1},\\
|C_f| &=& \sqrt{\frac{m_{f1} \, m_{f2} \, m_{f3}}{|A_f|}} \label{parametro3c1}.
\end{eqnarray}
Therefore, the value of parameter $|A_f|$ depends on the parity of $m$: for
$m$ even we have $m_{f1} \leq|A_f|\leq m_{f1} -m_{f2} +m_{f3}$, whereas $m$ odd
leads to $m_{f1} -m_{f2} +m_{f3} \leq|A_f|\leq m_{f3}$. We shall assume a
linear behavior of $|A_f|$ in terms of the parameters $0 \leq \beta_i^f
\leq 1$ ($i=1,2$), so that one can write
$|A_f|=m_{f1} \left(1+ \beta_1^f \, \frac{m_{f3} - m_{f2}}{m_{f1}} \right)$, for $m$
even, and $|A_f|= m_{f3} \left(1 -\beta_2^f \, \frac{m_{f2} - m_{f1}}{m_{f3}} \right)$,
for $m$ odd. The idea is then to expand $|A_f|$ in terms of 
$z=\frac{m_{f3} - m_{f2}}{m_{f1}}$, for $m$
even, and $z=\frac{m_{f2} - m_{f1}}{m_{f3}}$, for $m$ odd.
Considering now a four-zero texture form for the mass matrix $M_f$ and choosing
$A_L^f=\textrm{Diag}\left[ 1,\frac{d_2^f}{c_2^f},\frac{b_2^{f*}}{c_2^f} \right]$ and
$A_R^f=\textrm{Diag}\left[ \frac{|c_2^f|^2}{d_2^f},c_2^f,\frac{b_2^f c_2^f}{d_2^f} \right]$,
we thus have
\begin{equation}
Y_{2}^f=\left(
\begin{array}{ccc}
 0 & c_2^f \, C_f & 0 \\
 c_2^{f*} \, C_f^* & d_2^f \widetilde{B}_f & b_2^f B_f \\
 0 & b_2^f B_2^{f*} & a_2^f A_f
\end{array}
\right),  \label{Y2-final}
\end{equation}
where $a_2^f =\frac{|b_2^f|^2}{d_2^f}$. The
Yukawa matrix preserves the four-zero texture form. Thus, for $m$ odd, 
$|A_f|= m_{f3} \left(1 -\beta_2^f \, \frac{m_{f2} - m_{f1}}{m_{f3}} \right)$, one can reproduce the parametrization of a four-zero texture Yukawa matrix  given in Ref.\cite{DiazCruz:2004tr,DiazCruz:2004pj}. So,
the Cheng-Sher {\it ansatz} from Eq.(\ref{eq2}) can be reproduced in the limit
$m_{f1} << m_{f2} << m_{f3}$, and the parameters $\widetilde{\chi}_{ij}^f$ can be
written in terms of the entries of $A_L^f$ and $A_R^f$ matrices,
\begin{eqnarray}
\widetilde{\chi}_{11}^f &=& \left[ d_2^f -(c_2^{f*} e^{i \phi_{cf}} +c_2^f e^{-i\phi_{cf}}),
\right] \eta^f + \left[a_2^f + d_2^f -(b_2^{f*} e^{i\theta_{bf}} +b_2^f e^{-i\theta_{bf}})\right]\beta_2^f,  \\
\widetilde{\chi}_{12}^f &=& c_2^f e^{-i\theta_{cf}} -d_2^f -\eta^f \left[a_2^f +d_2^f
-(b_2^{f*} e^{i\theta_{Bf}} +b_2^f e^{-i\theta_{Bf}})\right] \beta_2^f, \\
\widetilde{\chi}_{13}^f &=& (a_2^f -b_2^f e^{-i\theta_{Bf} })\eta^f \sqrt{\beta_2^f}, \\
\widetilde{\chi}_{22}^f &=& d_2^f \eta^f +\left[a_2^f +d_2^f -(b_2^{f*} e^{i\theta_{Bf} }
+b_2^f e^{-i\theta_{Bf}})\right]\beta_2^f, \\
\widetilde{\chi}_{23}^f &=& (b_2^f e^{-i\theta_{Bf}} -a_2^f) \sqrt{\beta_2^f}, \\
\widetilde{\chi}_{33}^f &=& a_2^f,
\end{eqnarray}
where $\eta^f=\lambda^f_2/m^f_2$, with $m^f_2=|\lambda^f_2|$, and $0 \leq \beta_2^f
\leq 1$. The above equations can be inverted
to estimate the entries of matrices $A_L$ and $A_R$. This case has been studied previously and the same constraints in the parameters $\chi_{ij}$ can be imposed. In particular, we find
that $\chi_{ij}^q= O(1)$ are allowed  \cite{DiazCruz:2004pj,DiazCruz:2004tr,GomezBock:2005hc,DiazCruz:2009ek,BarradasGuevara:2010xs}.  In figure \ref{yukawa} are shown the values of the Yukawa
matrix entries when $|\widetilde\chi_{ij}|\sim 1$ and the phases are taken to be zero. The sudden fall and rise of $|(Y^f_{2})_{12}|$ in the upper right plot of Fig.1 stems from a sign change in the Yukawa coupling value. This case represents a special approximation when $\textrm{Arg}(C^f_a)=\textrm{Arg}(\widetilde{C}^f_a)$ (see table \ref{versions2HDM}).

 On the other hand, for $m$ even  $|A_f|=m_{f1} \left(1+ \beta_1^f \, \frac{m_{f3} - m_{f2}}{m_{f1}} \right)$, in which case, one can see from Eq. (\ref{Y2-final})  that the Yukawa coupling form changes and therefore  its parametrization. Although the analytical expressions of Yukawa texture for m-even  are larger than for the m-odd case, in general, we get that $(Y_{2}^f)_{\small m-even} \propto ( Y_{2}^f)_{\small m-odd}$. We shall discuss in in detail the phenomenology of this scenario in a forthcoming paper, however, for practical reasons,  this case is here analyzed numerically.
For the quark sector
it is possible to estimate the value
of $\beta_1^u$, $\beta_1^d$ by using the experimental information of the CKM matrix.
Our analysis gives $\beta_1^u = \beta_1^d \sim 0.9985$, to be compared with
the Yukawa couplings including the Cheng-Sher parametrization.
For the up-type quark sector we have
\begin{equation}
\frac{|(Y^{even,u}_{2})_{11}|}{|(Y^{CS,u}_{2})_{11}|}\sim 13.08; \qquad
\frac{|(Y^{even,u}_{2})_{12}|}{|(Y^{CS,u}_{2})_{12}|}\sim 6.85; \qquad
\frac{|(Y^{even,u}_{2})_{13}|}{|(Y^{CS,u}_{2})_{13}|}\sim 8.78,
\end{equation}
\begin{equation}
\frac{|(Y^{even,u}_{2})_{22}|}{|(Y^{CS,u}_{2})_{22}|}\sim 3.56; \qquad
\frac{|(Y^{even,u}_{2})_{23}|}{|(Y^{CS,u}_{2})_{23}|}\sim 5.00; \qquad
\frac{|(Y^{even,u}_{2})_{33}|}{|(Y^{CS,u}_{2})_{33}|}\sim 0.80.
\end{equation}
In the down-type quark sector,
\begin{eqnarray}
\frac{|(Y^{even,d}_{2})_{11}|}{|(Y^{CS,d}_{2})_{11}|}\sim 8.38; \qquad
\frac{|(Y^{even,d}_{2})_{12}|}{|(Y^{CS,d}_{2})_{12}|}\sim 3.69; \qquad
\frac{|(Y^{even,d}_{2})_{13}|}{|(Y^{CS,d}_{2})_{13}|}\sim 5.13,
\end{eqnarray}
\begin{eqnarray}
\frac{|(Y^{even,d}_{2})_{22}|}{|(Y^{CS,d}_{2})_{22}|}\sim 1.57; \qquad
\frac{|(Y^{even,d}_{2})_{23}|}{|(Y^{CS,d}_{2})_{23}|}\sim 2.64; \qquad
\frac{|(Y^{even,d}_{2})_{33}|}{|(Y^{CS,d}_{2})_{33}|}\sim 0.69.
\end{eqnarray}
and FCNC are under control. For  leptons ($\ell$) we obtain for all cases that
\begin{eqnarray}
\frac{|(Y^{even,\ell}_2)_{ij}|}{|(Y^{CS,\ell}_2)_{ij}|}\sim O(1).
\end{eqnarray}
Based on these results and following Eqs. (\ref{gen-Y}-\ref{gen-chi}), one can obtain for all fermions
\begin{eqnarray}
 \frac{|\widetilde\chi^{even,f}_{ij}|}{|\widetilde\chi^{CS,f}_{ij}|}= \frac{|(Y^{even,f}_2)_{ij}|}{|(Y^{CS,f}_2)_{ij}|}.
\end{eqnarray}
We have thus managed to implement  all experimental constraints found previously \cite{DiazCruz:2004pj,DiazCruz:2004tr,GomezBock:2005hc,DiazCruz:2009ek,BarradasGuevara:2010xs}. In this report we find strong constraints for the free parameters $\widetilde\chi^{even,f}_{ij}$. Following References \cite{DiazCruz:2004pj,DiazCruz:2004tr},  we can obtain the constraint
$|\widetilde \chi^{even,\ell}_{12}| \leq 5 \times10^{-1}$
from $\mu^- -e^-$ conversion,
$|\widetilde\chi^{even,\ell}_{13}|=|\widetilde\chi^{even,\ell}_{23}| \leq  10^{-2}$ from radiative decay $\mu^+ \to e^+ \gamma$,  and
$|\widetilde\chi^{even,d}_{23}| \leq 0.2 $ from the contribution to the decay $b \to s \gamma$ measurements. In addition, without loss of generality, we can implement all previous studies given in references \cite{DiazCruz:2004pj,DiazCruz:2004tr,GomezBock:2005hc,DiazCruz:2009ek,BarradasGuevara:2010xs} and we can validate the PA-2HDM as a framework phenomenologically viable, as well as the corresponding predictions.

\begin{acknowledgments}
This work has been supported in part by {\it Red de F\'{\i}sica de
Altas Energ\'{\i}as}, {\it Sistema Nacional de
Investigadores (CONACYT-M\'exico)} and also by \textit{PROMEP (M\'exico)}. J. H.-S. thanks A. Akeroyd for useful discussions during the Third International Workshop on
Prospects for Charged Higgs Discovery at Colliders
Uppsala University, Sweden, 27-30 September 2010.  The authors thank Dr. A. Flores-Riveros for
a careful reading of the manuscript.
\end{acknowledgments}

\end{document}